\documentclass[a4paper,10pt]{article}
\usepackage{graphicx}
\usepackage[numbers,sort&compress]{natbib}
\usepackage{url}
\usepackage{amsmath}
\usepackage{nicefrac}
\usepackage{amsfonts}
\usepackage{amssymb}
\usepackage{psfrag}
\usepackage{subfigure}
\usepackage{subeqnarray}
\usepackage{multicol}

\newcommand{\p}{\partial}

\newcommand{\n}{\nabla}

\newcommand{\ud}{\mathrm{d}}
\newcommand{\bld}{\boldsymbol}

\newcommand{\be}{\begin{equation}}
\newcommand{\ee}{\end{equation}}
\newcommand{\bn}{\boldsymbol{\nabla}}
\newcommand{\bx}{\boldsymbol{x}}
\newcommand{\bX}{\boldsymbol{X}}
\newcommand{\bv}{\boldsymbol{v}}

\newcommand{\bK}{\boldsymbol{K}}
\newcommand{\bM}{\boldsymbol{M}}

\newcommand{\bnn}{\boldsymbol{n}}

\newcommand{\bo}{\boldsymbol{0}}

\newcommand{\Ti}{\textnormal{\tiny \emph{T}}}
\newcommand{\Ii}{\textnormal{\tiny \emph{I}}}
\newcommand{\Si}{\textnormal{\tiny \emph{S}}}
\newcommand{\Gi}{\textnormal{\tiny \emph{$\Gamma$}}}
\newcommand{\T}{\mathsf{T}}

\begin{document}

\title{A simplified multiphase multiscale model for tissue growth}
\author{E.C. Holden$^1$, B.S. Brook$^1$, S.J. Chapman$^2$ \& R.D. O'Dea$^1$\\
              $^1$Centre for Mathematical Medicine and Biology,\\ School of Mathematical Sciences,\\
	      University of Nottingham, University Park, \\Nottingham, NG7 2RD, UK\\
	      $^2$Mathematical Institute, University of Oxford, Radcliffe Observatory Quarter,\\ 
	      Woodstock Road, Oxford OX2 6GG, UK
}



\maketitle

\begin{abstract}
In this paper, we derive an effective macroscale description suitable to describe the growth of biological tissue within a porous tissue-engineering scaffold. As in our recent work (Holden \textit{et al.} ``A multiphase multiscale model for nutrient limited tissue growth'', The ANZIAM Journal, 2018, doi:10.1017/S1446181118000044) the underlying tissue dynamics is described as a multiphase mixture, thereby naturally accommodating features such as interstitial growth and active cell motion. Via a linearisation of the underlying multiphase model (whose nonlinearity poses significant challenge for such analyses), we obtain, by means of multiple-scales homogenisation, a simplified macroscale model that nevertheless retains explicit dependence on both the microscale scaffold structure and the tissue dynamics. The model we obtain comprises Darcy flow, and differential equations for the volume fraction of cells within the scaffold and the concentration of nutrient, required for growth. These are coupled to underlying Stokes-type cell problems that provide permeability tensors to parameterise the macroscale description.  In Holden \textit{et al.}, the cell problems retain macroscale dependence, posing significant computational challenges; here, we obtain a decoupled system whereby the quasi-steady cell-problems may be solved separately from the macroscale description, thereby greatly reducing the complexity associated with fully-coupled multiscale descriptions. Moreover, we indicate how the formulation is influenced by a set of alternative microscale boundary conditions.\\
\end{abstract}

\section{Introduction}\label{intro}

Tissue growth is a complex and inherently multiscale phenomenon, whose unified description requires the integration of insight obtained at one scale with observations at another. For example growth processes (or disease manifestation) at the organ scale are driven by microscopic events at the (sub-)cellular scale that themselves are influenced by macroscopic dynamics. Such complexity necessitates investigation by theoretical means, to provide supporting insight not available by experimental investigation alone. However, the complex multiscale interactions (from intra- and inter-cell signalling pathways, cell biomechanics and migration to tissue-level patterning and mechanics) leads inevitably to formulations that are analytically and computationally intractable, or are otherwise highly idealised. For this reason, a significant area of research is dedicated to developing various mathematical and computational techniques that enable efficient coupling between dynamics occurring on multiple scales (see, \textit{e.g.} \cite{alarcon2006multiscale,macklin2009multiscale,osborne2010hybrid} and references therein).

This article is concerned with the method of multiple-scale asymptotic homogenisation that provides a coarse-scale description of the tissue dynamics, while still incorporating aspects of the microscale physics. Such methods have long theoretical history (e.g.~\cite{bensoussan1978asymptotic,keller1980darcy}), and have been widely used to describe descriptions of porous and poroelastic materials in applied/industrial settings, such as the study of soil and reservoirs \cite{popov2009multiscale,ptashnyk2010derivation}; see \cite{pavliotis2008multiscale,davit2013homogenization} for reviews. The key feature of this approach is to derive suitable macroscale equations from an underlying microscale description, rather than stating them \textit{ab initio}. Coupling to the microscale physics is effected by suitable problems defined on a prototypical `unit cell'. These so-called `cell problems', determine microscale behaviour that is subsequently employed to specify effective coefficients in the macroscale description.

More contemporary studies have employed these methods in a biological setting. Of particular relevance to the current work is a series of studies that seek to describe growing tissues. In \cite{odea2015multiscale} a simple solid-accretion-based model of nutrient-limited tissue growth within a porous scaffold, was considered. A macroscale description of growth and transport was obtained using a multiple-scales technique, to accommodate explicit dependence on microscale dynamics and structure. A similar analysis by Penta et al. \cite{penta2014effective} described accretion in a poroelastic setting. To permit analysis, the authors of \cite{odea2015multiscale,penta2014effective} (and other similar studies) exploit asymptotic restrictions on the underlying model, considering slow (quasi-static) growth and linearised deformation. In Collis et al. \cite{collis2017effective}, such assumptions are relaxed to consider a macroscale representation of finite volumetric nutrient-limited growth of a hyperelastic solid, employing the Arbitrary-Lagrangian-Eulerian approach \cite{brown2014effective}. Collis \textit{et al.} \cite{collis2016transport,collis2016numeric} sought to address the highly idealised representation of growth in the aforementioned studies, by employing a multiphase description for the underlying tissue dynamics that naturally   accommodates the complexity associated with tissue growth dynamics, such as interstitial growth and active cell motion. However, the simplifying adoption of a large-drag limit employed therein constrains growth to a thin boundary layer and the resulting model is effectively equivalent to accretion. This deficiency was addressed in Holden \textit{et al.} \cite{holden2018multiphase} to obtain an effective macroscale description of nutrient limited tissue growth on an artificial scaffold, in which the assumption of large interphase drag is relaxed so that active cell motion is permitted, caused by the cells' tendency to aggregate or repel. Analytical progress was effected by a linearisation that ameliorates problems associated with complex mass-transfer considered in the multiphase model (see \cite{collis2017effective} for a discussion), and allows one to obtain a more tractable description that permits coupling between micro- and macro-scale processes. The derived model comprises a Darcy flow, a partial differential equation for the volume fraction of cells within the scaffold, and an advection-reaction equation for the nutrient concentration, coupled to the underlying microscale dynamics via suitable cell problems. Importantly, and in contrast to other similar studies, these unit cell problems are themselves parameterised by the macroscale dynamics, so that the micro- and macro- scale descriptions are fully coupled. Here, we show how through a modification to the analysis of \cite{holden2018multiphase}, the more standard de-coupling between microscale and macroscale can be effected, leading to a system whereby the quasi-steady cell-problems may be solved separately from the macroscale description, thereby greatly simplifying the computational difficulty associated with fully-coupled multiscale descriptions. In addition, we indicate how this formulation is modified under alternative boundary conditions to those employed in \cite{holden2018multiphase}.

\section{Model formulation}\label{sec:model}

We consider a model of broad relevance to tissue engineering applications where tissue growth occurs on a structured periodic scaffold, such as can be achieved through the use of 3D printing \cite{jakus2015three,visser2015reinforcement}. We emphasise that the model set-up considered herein is identical to that of \cite{holden2018multiphase} (itself following closely \cite{lemon2006}, that builds on the general theory of multiphase porous flow developed in \cite{marle1982macroscopic,bowen1980incompressible,drew2006theory}), and so only a very brief summary is included here.

\begin{figure}
 \centering
 \includegraphics[width=.3\textwidth]{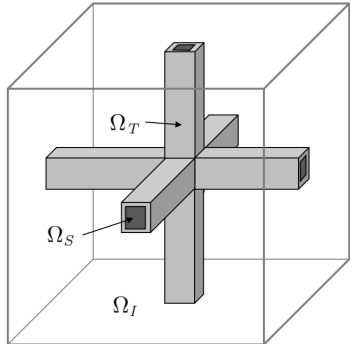}
 \caption{Schematic diagram of the microscale domain $\Omega$ illustrating a periodic scaffold covered with a layer of tissue, indicating the scaffold, $\Omega_{\Si}$, tissue $\Omega_{\Ti}$ and interstitial fluid $\Omega_{\Ii}$ domains. The scaffold boundary is denoted by $\partial \Omega_\Si$ and tissue-interstitial boundary by $\Gamma$.}
 \label{fig:dig}
\end{figure}

The microscale domain is denoted $\Omega$, with boundary $\partial \Omega$, and has characteristic lengthscale $l^*$. This domain comprises scaffold, $\Omega_{\Si}$, tissue $\Omega_{\Ti}$ and interstitial fluid $\Omega_{\Ii}$. The scaffold boundary is denoted by $\partial \Omega_{\Si}$ and the tissue-interstitial boundary by $\Gamma$; see Figure \ref{fig:dig} for a schematic diagram. The macroscopic lengthscale (associated with the full extent of the scaffold) is denoted $L$. The lengthscales in question are well-separated such that
\[
 \varepsilon= \frac{l}{L}\ll 1.
\]

We model the porous scaffold material as a rigid solid, and the tissue as a two phase mixture of cells and interstitial fluid which covers the scaffold, whilst the interstitial space contains only fluid. Henceforth, we refer to the interstitial fluid as water, for concision. Both cells and water are modelled as viscous fluids, described by a Stokes flow. Increase in the cell volume fraction of the mixture depends on the concentration of a generic diffusible nutrient, as well as the availability of water. Tissue growth is represented by movement of the boundary $\Gamma$, occuring as a consequence of nutrient limited phase transition or cell aggregation/repulsion.

\subsection{Model equations}

The equations governing the multiphase mixture in the tissue domain are as follows:
\begin{align}
  &\theta_n+\theta_w =1,\\
  &\rho_{i}\left( \frac{\partial \theta_{i}}{\partial t} + \boldsymbol{\n} \cdot \left(\theta_{i}\boldsymbol{v}_{i}\right)\right)=S_{i};\ i=n,w,\\
  &\boldsymbol{\n} \cdot \left(\theta_n\boldsymbol{v}_n+\theta_w\boldsymbol{v}_w\right) = \left(\frac{1}{\rho_n}-\frac{1}{\rho_w}\right)S_n,\label{eq:masscons}\\
  &\boldsymbol{\n} \cdot \left(\theta_{i}\boldsymbol{\sigma}_{i}\right) + \boldsymbol{f}_{ij}=\boldsymbol{0};\ i=n,w,\label{eq:mom}
\end{align}
wherein the subscript `$n$' or `$w$' denotes variables associated with the cell or water phases, respectively. The volume fraction of the $i^\textrm{th}$ phase is denoted $\theta_{i}$, with associated density $\rho_{i}$, velocity $\boldsymbol{v}_{i}$, mass source $S_{i}$ (obeying $S_n=-S_w$). The stress tensor of phase $i$ is denoted $\boldsymbol{\sigma}_{\Ii}$ and the interphase force associated with the action of phase $j$ on phase $i$ by $\boldsymbol{f}_{ij}$, these being defined as follows:
\begin{align}
  &\boldsymbol{\sigma}_{i} = -p_{i}{\boldmath{I}}+\mu_{i}\left( \boldsymbol{\n}\boldsymbol{v}_{i} + \left( \boldsymbol{\n}\boldsymbol{v}_{i}\right)^\top - \frac{2}{3}\boldsymbol{\n}\cdot\boldsymbol{v}_{i} \boldmath{I} \right),\label{eq:sigma}\\
  &\boldsymbol{f}_{ij} = p\boldsymbol{\n}\theta_{i} + \beta\theta_{i} \theta_j \left( \boldsymbol{v}_j-\boldsymbol{v}_{\Ii}\right).\label{eq:fij}
\end{align}
In (\ref{eq:sigma}) $p_{i}$ denotes the pressure of the $i^\textrm{th}$ phase, whose constant viscosity is $\mu_{i}$. Consistent with the two-phase representation we adopt, in (\ref{eq:fij}) we have assumed that the only interphase interaction that exists is passive viscous drag, with coefficient $\beta$. The common `mixture pressure' is denoted by $p$, which is related to the individual phase pressures by $p_w= p$ and $p_n = p +\phi_n$, where $\phi_n$ represents an additional intraphase pressure, generated by cell-cell interactions, that can lead to active cell motion. This is specified as:
\begin{equation}
	\phi_n = \theta_n\left(-\nu + \frac{\kappa \theta_n}{1-\theta_n}\right),\label{eq:phi_n}
\end{equation}
in which the first term represents aggregation with strength $\nu>0$, while the second term curtails this, with strength $\kappa>0$. Note in particular the singularity that occurs as $\theta_n\to 1$, representing high repulsion occurring when all available space is occupied by cells.  Lastly, we note that it is at times notationally convenient to refer to weighted mixture variables, represented by the subscript $T$; for example, we define: $\boldsymbol{v}_{\Ti} = \theta_n \boldsymbol{v}_n + \theta_w\boldsymbol{v}_w$. We reiterate that the above described model is identical to that presented in \cite{holden2018multiphase} and follows closely that developed in \cite{lemon2006}.

As noted above, in the interstitium ($\Omega_\Ii$), a corresponding viscous flow model is adopted; however, since there are no cells ($\theta_n=0$, $\theta_w=1$) equations (\ref{eq:masscons})--(\ref{eq:fij}) reduce to a standard incompressible Stokes flow. Variables in this domain are denoted with a subscript $I$.

A generic nutrient of concentration $c$, on which mitosis depends, is advected by the flows, diffuses (with diffusivity $D_{\Ii}$), and in the tissue domain $\Omega_{\Ti}$ is taken up by the cell phase at a rate $\Lambda$ according to:
\begin{align}
	& \frac{\p c_{\Ii}}{\p t}+\boldsymbol{\n}\cdot \left(c_{\Ii} \boldsymbol{v}_{\Ii}\right)=\boldsymbol{\n}\cdot\left(D_{\Ii}\boldsymbol{\n}c_{\Ii}\right),\ \mbox{in}\ \Omega_{\Ii},\\
		& \frac{\p c_{\Ti}}{\p t}+\boldsymbol{\n}\cdot \left(c_{\Ti} \boldsymbol{v}_{\Ti}\right)=\boldsymbol{\n}\cdot\left(D_{\Ti}\boldsymbol{\n}c_{\Ti}\right)-\Lambda,\ \mbox{in}\ \Omega_{\Ti}\label{eq:cT}.
\end{align}
Equation (\ref{eq:cT}) arises from the sum of the nutrient transport equations in each phase, under the assumption that due to rapid equilibriation across cell membranes, the concentration in the cell and water phases is equal (as employed in \cite{lemon2007multiphase}). We assume that the diffusivity in the interstitium, $D_{\Ii}$ is constant, but specify $D_{\Ti}=D_{\Ti}(\theta_n)$, and $\Lambda=\Lambda(\theta_n,c)$.

\subsection{Boundary conditions}

At the scaffold-tissue boundary, we impose no-slip and no-penetration conditions, so that:
\begin{align}
	\boldsymbol{v}_{i}=\boldsymbol{0}, \quad \boldsymbol{\n}c\cdot \boldsymbol{n}_{\Si}=0\ \mbox{on}\ \Omega_{\Si},\label{eq:BCnoslip}
\end{align}
wherein $\boldsymbol{n}_{\Si}$ denotes the outward normal to the scaffold surface.

The tissue and interstitial domains are separated by the free interface $\Gamma$, whose evolution is given by
\begin{equation}
	\frac{\p F}{\p t}+\boldsymbol{\n}F\cdot \boldsymbol{v}_{\Gi} = 0,
\end{equation}
where $\boldsymbol{v}_{\Gi}$ is the boundary velocity and $F$ denotes the position of the moving interface through the level set equation $F(x,t)=0$. On this interface, the dynamics are coupled via the following flux and stress continuity conditions:
\begin{align}
	&\rho_n\theta_n\left(\boldsymbol{v}_n-\boldsymbol{v}_{\Gi}\right)\cdot \boldsymbol{n}=0,\label{eq:BCflux1}\\
	&\rho_w\theta_w\left(\boldsymbol{v}_w-\boldsymbol{v}_{\Gi}\right)\cdot \boldsymbol{n}=\rho_w\theta_w\left(\boldsymbol{v}_{\Ii}-\boldsymbol{v}_{\Gi}\right)\cdot \boldsymbol{n},\label{eq:BCflux2}\\
	&\left[ c_{i} \left( \boldsymbol{v}_{i}-\boldsymbol{v}_{\Gi} \right)\cdot \boldsymbol{n} - D_{i}\boldsymbol{\n}c_{i}\cdot\boldsymbol{n} \right]^+_-=0\label{eq:BCstress1}\\
	&\left[ \boldsymbol{v}_{i}\cdot\boldsymbol{t} \right]^+_-=0,\ 	\left[ \boldsymbol{\sigma}_{i}\cdot\boldsymbol{n} \right]^+_-=0,\quad 	\left[ c_{i} \right]^+_-=0,\label{eq:BCstress2}
\end{align}
where $i=T,I$, $\boldsymbol{n}$ is the outward normal to $\Gamma$ and $[\ ]^+_-$ denotes the jump across the interface. We remark that our model describes a complex free-boundary problem in which the interface position $\Gamma$ is not known, and should be determined as part of the solution. However, in the multiscale analysis that follows, the boundary velocity remains undetermined. In order to close the model, we are therefore required to specify constitutively this motion; this issue is considered in more detail in \cite{holden2018multiphase}.

\subsection{Nondimensionalisation and linearisation}

We non-dimensionalise our model equations by using the following scalings
\begin{align}
	\boldsymbol{x}=l\hat{\boldsymbol{x}},\ 
	\boldsymbol{v}_{i}=V\hat{\boldsymbol{v}}_{i},\
	c_{i} = C \hat{c}_{i},\
	p = \frac{\mu_nV}{l}\hat{p},\
	t = \frac{l}{V}\hat{t},\\
	S_n=\frac{\rho_nV}{l}\hat{S}_n,\
	\beta = \frac{\mu_n}{l^2}\hat{\beta},\
	\frac{D_{i}}{Vl} = \frac{1}{\textrm{Pe}_{i}},\
	\Lambda = \frac{V}{l}\hat{\Lambda}
\end{align}
in which circumflexes denote dimensionless variables, $i=n,w,T,I$, and $V$ and $C$ are a characteristic microscale velocity and nutrient concentration. Henceforth, we drop the circumflex notation for simplicity.

We reduce the degree of nonlinearity of the microscale model to enable a more straightforward multiscale analysis by linearising the equations about a uniform steady state, across $\Omega_{\Ti}$, as follows:
\begin{equation}
	\theta_n = \theta^*_n + \delta \theta_{n,1}+\cdots,
\end{equation}
with corresponding expansions for the other model variables, and where $0<\delta\ll 1$ and asterisks denote steady-state values. For concision we do not state the linearised model here (the reader is referred to equations (2.27)--(2.32) in \cite{holden2018multiphase}) but we highlight  in particular that the steady state volume fraction, nutrient concentration and velocity are defined by
\begin{equation}
	S_n(\theta_n^*,\theta_w^*,c^*)=0,\ \Lambda(\theta_n^*,c^*)=0,\label{eq:steadystate}
\end{equation}
and $\boldsymbol{v}_{i}^*=\boldsymbol{0}$. Moreover, the source, uptake and intraphase interaction terms that appear in the following are the first order linear corrections, defined by:
\begin{align}
  &S_{n,1} = \frac{\p S_n}{\p \theta_n}\left(\theta_n^*,c^*\right)\theta_{n,1} + \frac{\p S_n}{\p c}\left(\theta_n^*,c^*\right)c_{\Ti,1},\quad
  \phi_{n,1} = \frac{1}{\theta_n^*}\frac{\p \left( \theta_n\phi_n\right)}{\p \theta_n}\left(\theta_n^*\right)\theta_{n,1},\label{eq:linfuncs1}\\
  &\Lambda_{1} = \frac{\p \Lambda}{\p \theta_n}\left(\theta_n^*,c^*\right)\theta_{n,1} + \frac{\p \Lambda}{\p c}\left(\theta_n^*,c^*\right)c_{\Ti,1}.\label{eq:linfuncs2}
\end{align}

\section{Multiple scales analysis}

We now work with the linearised version of the model described in \S\ref{sec:model} and, for the sake of clarity, drop the associated subscripts. 

To derive a suitable macroscale description incorporating the microscale growth, dynamics and structure, we follow (\textit{e.g.}) \cite{collis2016numeric,odea2015multiscale,shipley2010multiscale} in using the method of multiple scales. Correspondingly we rescale such that the timescale under consideration is that of macroscale advection and the pressure scaling results in the appropriate leading order problem:
\begin{equation}
	t=\varepsilon \tilde{t},\  p = \frac{1}{\varepsilon}\tilde{p}
\end{equation}
in which tildes denote the rescaled variables. We drop the tilde notation for convenience as we work exclusively with the rescaled variables in subsequent sections. This choice of time rescaling simplifies the analysis by resulting in a quasi-steady problem at leading order.

Next we introduce a macroscale coordinate $\boldsymbol{X}$ where $\boldsymbol{X}=\varepsilon \boldsymbol{x}$ ($\boldsymbol{x}$ being the microscale coordinate) and expand in multiple-scales form as follows:
\begin{align}
	&\psi(\boldsymbol{x},\boldsymbol{X},t;\varepsilon)=\psi^{(0)}(\boldsymbol{x},\boldsymbol{X},t)+\varepsilon \psi^{(1)}(\boldsymbol{x},\boldsymbol{X},t) + \ldots\\
	&\boldsymbol{\n}=\boldsymbol{\n}_x+\varepsilon \boldsymbol{\n}_X,\ 	{\n}^2={\n}^2_x+2\varepsilon \boldsymbol{\n}_x\cdot \boldsymbol{\n}_X+\varepsilon^2 \n^2_X.
\end{align}
Moreover, in addition to the boundary conditions (\ref{eq:BCnoslip}), (\ref{eq:BCflux1})--(\ref{eq:BCstress2}) we require that $\psi^{(i)}$ for $i= 0, 1, \ldots$ are periodic in $\boldsymbol{x}$. We now analyse the equations at each order in $\varepsilon$, with the aim of obtaining a description of the macroscale growth and transport

\subsection{Microscale governing equations at each order in $\varepsilon$}\label{sec:eqsateachorder}

At O$(1)$, the equations and boundary conditions in the tissue domain $\Omega_\Ti$ are as follows:
\begin{align}
\theta_n^*\bn_{\bx}\cdot\bv_{n}^{(0)}&=S_{n}^{(0)}\label{eqn:3ddndtO1},\\
\theta_w^*\bn_{\bx}\cdot\bv_{w}^{(0)}&=-\bar{\rho}S_{n}^{(0)}\label{eqn:3ddwdtO1},\\
\theta_n^{(0)}+\theta_w^{(0)}&=0\label{eqn:novoidsO1},\\
\bn_{\bx}\cdot\bv_{\Ti}^{(0)}&=\left(1-\bar{\rho}\right)S_{n}^{(0)},\label{eqn:3ddivO1}\\
\theta_n^*\bn_{\bx} \left(p_{w}^{(0)} + \phi_{n}^{(0)}\right)&=\boldsymbol{0},\label{eqn:3dpnO1}\\
\theta_w^*\bn_{\bx} p_{w}^{(0)}&=\boldsymbol{0},\label{eqn:3dpwO1}\\
\bn_{\bx}\cdot\left(c^*\bv^{(0)}_{\Ti}\right)&=\frac{1}{Pe_\Ti}\nabla^2_{\bx} c_{\Ti}^{(0)}-\Lambda^{(0)}.
\end{align}
In the interstitial domain, $\Omega_\Ii$:
\begin{align}
\bn_{\bx}\cdot\bv_{\Ii}^{(0)}&=0,\label{eqn:O1I1}\\
\bn_{\bx}p_{\Ii}^{(0)}&=\boldsymbol{0},\label{eqn:pI}\\
\bn_{\bx}\cdot\left(c^*\bv_{\Ii}^{(0)}\right)&=\frac{1}{Pe_\Ii}\nabla^2_{\bx} c_{\Ii}^{(0)}.
\end{align}
On the interface, $\Gamma$, the boundary conditions are:
\begin{align}
\theta_n^*\left(\bv_{n}^{(0)}-\bv_{\Gi}^{(0)}\right)\cdot\bnn&=0,\label{eqn:O1bc1}\\
\theta_w^*\left(\bv_{w}^{(0)}-\bv_{\Gi}^{(0)}\right)\cdot\bnn&=\left(\bv_{\Ii}^{(0)}-\bv_{\Gi}^{(0)}\right)\cdot\bnn,\label{eqn:O1bc2}\\
\bv_{\Ti}^{(0)}\cdot\boldsymbol{t}&=\bv_{\Ii}^{(0)}\cdot\boldsymbol{t},\label{eqn:O1t}\\
-\left(p_{w}^{(0)}+\theta_n^*\phi_{n}^{(0)}\right)\boldsymbol{I}&=-p_{\Ii}^{(0)}\boldsymbol{I},\label{eqn:bcstressO1}\\
\bn_{\bx} F^*\cdot\bv_{\Gi}^{(0)}&=0,\label{eqn:bcevo}\\
c^* \left(\bv^{(0)}_{\Ti}-\bv_{\Gi}^{(0)}\right)\cdot\bnn-\frac{1}{Pe_\Ti}\bn_{\bx} c_{\Ti}^{(0)}\cdot\bnn&=c^* \left(\bv_{\Ii}^{(0)}-\bv_{\Gi }^{(0)}\right)\cdot\bnn-\frac{1}{Pe_\Ii}\bn_{\bx} c_{\Ii}^{(0)}\cdot\bnn,\\
c_\Ti^{(0)}&=c_\Ii^{(0)},
\end{align}
where
\be
\bv_{\Ti}^{(0)}=\theta_n^*\bv^{(0)}_{n}+\theta_w^*\bv^{(0)}_{w}.
\ee
On the scaffold surface, $\partial\Omega_\Si$, we impose:
\be\label{eqn:O1bcS}
\bv_{n}^{(0)}=\bv_{w}^{(0)}=\bo, \quad \bn_{\bx}c_{\Ti}^{(0)}\cdot\bnn_\Si=0.
\ee
Lastly, we note that, in view of (\ref{eq:linfuncs1}) and (\ref{eq:linfuncs2}), the phase transfer, intraphase pressure and nutrient uptake functions depend only on $\theta_n^{(0)}$, $c_\Ti^{(0)}$ (and the relevant steady states), so that
\be\label{eqns:formfuncsLO}
S_n^{(0)}=S_{n,1}\left(\theta_n^{(0)},c_\Ti^{(0)}\right),\quad \phi_n^{(0)}=\phi_{n,1}\left(\theta_n^{(0)}\right),\quad \Lambda^{(0)}=\Lambda_{1}\left(\theta_n^{(0)},c_\Ti^{(0)}\right).
\ee

Equation (\ref{eqn:bcevo}) tells us that $\bn_{\bx} F^*=\bo$ or $\bv_{\Gi}^{(0)}\cdot\bnn=0$, but the latter holds most generally, so we take the boundary to be stationary at this order and, for consistency, we rescale $S_{n}^{(0)}$ and $\Lambda^{(0)}$ to $O(\varepsilon)$. Following the arguments in \cite{odea2015multiscale,collis2016transport,holden2018multiphase} we find that pressures, nutrient concentrations and cell volume fraction are independent of the microscale variable $\bx$, i.e. 
\begin{align}
&p^{(0)}(\bX,t)=p_{\Ii}^{(0)}(\bX,t)=p_{w}^{(0)}(\bX,t)+\theta_n^*\phi_{n}^{(0)}(\bX,t),\label{O1paremacroscale}\\
&c^{(0)}(\bX,t)=c_{\Ti}^{(0)}(\bX,t)=c_{\Ii}^{(0)}(\bX,t),\\
&\theta_n^{(0)}=\theta_n^{(0)}(\bX,t).
\end{align}

At O($\varepsilon$) the governing equations in $\Omega_\Ti$ are:
\begin{gather}
\frac{\partial \theta_n^{(0)}}{\partial t}+\theta_n^*\left(\bn_{\bx}\cdot\bv_{n}^{(1)}+\bn_{\bX}\cdot\bv_{n}^{(0)}\right)=S^{(0)}_{n},\label{eqn:3ddndtO2}\\
\bn_{\bx}\cdot\bv_{\Ti}^{(1)}+\bn_{\bX}\cdot\bv_{\Ti}^{(0)}=\left(1-\bar{\rho}\right)S^{(0)}_{n},\label{eqn:oeT2}\\
\theta_n^*\left[\bn_{\bx} \left(p_{w}^{(1)}+\phi_{n}^{(1)}\right)+\bn_{\bX}\left( p_{w}^{(0)} +\phi_{n}^{(0)}\right)+\beta\theta_w^*\left(\bv_{n}^{(0)}-\bv_{w}^{(0)}\right)-\nabla^2_{\bx}\bv_{n}^{(0)}\right]=\boldsymbol{0},\label{eqn:3dpnO2}\\
\theta_w^*\left[\bn_{\bx} p_{w}^{(1)}+ \bn_{\bX} p_{w}^{(0)}+\beta \theta_n^*\left(\bv_{w}^{(0)}-\bv_{n}^{(0)}\right)-\mu\nabla^2_{\bx}\bv_{w}^{(0)}\right]=\boldsymbol{0},
\label{eqn:3dpwO2}\\
\frac{\partial c^{(0)}}{\partial t}+c^*\left(\bn_{\bx}\cdot\bv_{\Ti}^{(1)}+\bn_{\bX}\cdot\bv_{\Ti}^{(0)}\right)=\frac{1}{Pe\Ti}\nabla^2_{\bx} c_{\Ti}^{(1)}-\Lambda^{(0)},\label{eqn:dcdt1}
\end{gather}
where $\bv_{\Ti}^{(1)}=\theta_n^*\bv_{n}^{(1)}+\theta_w^*\bv_{w}^{(1)}$.
In $\Omega_\Ii$:
\begin{align}
\bn_{\bx}\cdot\bv_{\Ii}^{(1)}+\bn_{\bX}\cdot\bv_{\Ii}^{(0)}&=0,\label{eqn:oeI1}\\
\bn_{\bx}p_{\Ii}^{(1)}+\bn_{\bX}p^{(0)}-\mu\nabla^2_{\bx}\bv_{\Ii}^{(0)}&=\boldsymbol{0},\label{eqn:OeI2}\\
\frac{\partial c^{(0)}}{\partial t}+c^*\left(\bn_{\bx}\cdot\bv_{\Ii}^{(1)}+\bn_{\bX}\cdot\bv_{\Ii}^{(0)}\right)&=\frac{1}{Pe_\Ii}\nabla^2_{\bx} c_{\Ii}^{(1)}.\label{eqn:dcdt2}
\end{align}
On the tissue-intertitium interface, $\Gamma$, we obtain:
\begin{align}
\theta_n^*\left(\bv_{n}^{(1)}-\bv_{\Gi}^{(1)}\right)\cdot\bnn&=0,\label{eqn:oebc1}\\
\theta_w^*\left(\bv_{w}^{(1)}-\bv_{\Gi}^{(1)}\right)\cdot\bnn&=\left(\bv_{\Ii}^{(1)}-\bv_{\Gi}^{(1)}\right)\cdot\bnn,\label{eqn:oebc2}\\
\bv_{\Ti}^{(1)}\cdot\boldsymbol{t}&=\bv_{\Ii}^{(1)}\cdot\boldsymbol{t},
\end{align}
\begin{multline}
-\left(p_{w}^{(1)}+\theta_n^*\phi_{n}^{(1)}\right)\boldsymbol{I}+ \theta_n^* \left(\bn_{\bx}\bv_{n}^{(0)}+\left(\bn_{\bx}\bv_{n}^{(0)}\right)^\T\right)\\
+\mu \theta_w^* \left(\bn_{\bx}\bv^{(0)}_{w}+\left(\bn_{\bx}\bv_{w}^{(0)}\right)^\T\right)=-p_{\Ii}^{(1)}\boldsymbol{I}+\mu \left(\bn_{\bx}\bv_{\Ii}^{(0)}+\left(\bn_{\bx}\bv_{\Ii}^{(0)}\right)^\T\right),\label{eqn:oebc4}
\end{multline}
\begin{gather}
\frac{\partial F^{(0)}}{\partial t} +\bn_{\bx} F^*\cdot\bv_{\Gi}^{(1)}=0,
\end{gather}
\begin{multline}
c^*\left(\bv_{\Ti}^{(1)}-\bv_{\Gi}^{(1)}\right)\cdot\bnn-\frac{1}{Pe_\Ti}\left(\bn_{\bx} c_{\Ti}^{(1)}+\bn_{\bX} c^{(0)}\right)\cdot\bnn\\
=c^*\left(\bv_{\Ii}^{(1)}-\bv_{\Gi}^{(1)}\right)\cdot\bnn-\frac{1}{Pe_\Ii}\left(\bn_{\bx} c_{\Ii}^{(1)}+\bn_{\bX} c^{(0)}\right)\cdot\bnn,\label{eqn:oedc}
\end{multline}
\begin{gather}
c_{\Ti}^{(1)}=c_{\Ii}^{(1)}.
\end{gather}
Finally, on $\partial\Omega_\Si$ we apply:
\be\label{eqn:bcos2c}
\bv_{n}^{(1)}=\bv_{w}^{(1)}=\bo,\quad \bn_{\bx}c_{\Ti}^{(1)}\cdot\bnn+ \bn_{\bX}c_{\Ti}^{(0)}\cdot\bnn_\Si=0.
\ee

\subsection{Macroscale description}

In this subsection, we seek a macroscale representation of the flow, growth and transport described by the above equations. To effect this, we require a method of averaging variables across the various domains of the periodic cell. We therefore define the following integral average for some variable, $g$, over domain $\Omega_{\Ii}$ by
\be
	\langle g \rangle = \frac{1}{|\Omega|}\int_{\Omega_{\Ii}} g\, \ud V,\label{eq:av}
\ee
where $\Omega=\Omega_{\Ii}\cup \Omega_{\Ti} \cup \Omega_{\Si}$.

\subsubsection{Velocity and pressure ansatz}


To determine the flow in $\Omega_\Ti$ and $\Omega_\Ii$ specified by the system of equations given in \S\ref{sec:eqsateachorder}, we follow, \textit{e.g.}, \cite{odea2015multiscale,collis2016transport,davit2013homogenization,shipley2010multiscale,mei1991effect} in exploiting the linearity of the momentum equations (\ref{eqn:3dpnO2}), (\ref{eqn:3dpwO2}) and (\ref{eqn:OeI2}) by taking an appropriate form for the macroscale velocities and microscale pressures to be given by the following ansatz:
\be
\bv_{i}^{(0)}=-\bK_{i} \bn_{\bX} p^{(0)}\mbox{ and } p_{i}^{(1)}=-\boldsymbol{a}_{i}\cdot \bn_{\bX} p^{(0)}-\bar{p}_{i}, \label{eqn:appform}
\ee
where $p^{(0)}$ is the overall macroscale pressure, $\bK_{i}$ are tensors describing the permeability, $\boldsymbol{a}_{\Ii}$ are first order tensors imparting microscale pressure variation, and $\bar{p}_{i}$ are the mean (microscale-invariant) values of the first order pressures in $\Omega_{\Ii}$.

In the linearised model of \cite{holden2018multiphase}, this choice of ansatz results in unit cell problem that are parameterised by the macroscale pressure and cell volume fraction (through $\phi^{(0)}$) so that the micro- and macro-scale descriptions are fully-coupled (see equations (3.12)--(3.14) in that paper). This provides a significant challenge from a computational point of view. Here, we seek to remove this complexity; since both the macroscale pressure and active cell behaviour terms appear linearly in the momentum equations (\ref{eqn:3dpnO2}), (\ref{eqn:3dpwO2}) and (\ref{eqn:OeI2}), a more appropriate for the macroscale velocities and microscale pressures takes the following form:
\begin{align}
&\bv_{i}^{(0)}=-\bK_{i} \bn_{\bX} p_w^{(0)}-\bM_{i} \bn_{\bX} \phi_n^{(0)},\label{eqn:newansatz1}\\
&p_{i}^{(1)}=-\boldsymbol{a}_{i}\cdot \bn_{\bX} p_w^{(0)}-\boldsymbol{b}_{i}\cdot \bn_{\bX} \phi_n^{(0)}-\bar{p}_{i}.\label{eqn:newansatz2}
\end{align}
Note that rather than an ansatz in terms the overall macroscale pressure $p^{(0)}$, in equations (\ref{eqn:newansatz1}), (\ref{eqn:newansatz2}) the macroscale velocities and microscale pressures are given as linear functions of the macroscale common mixture pressure $p_w^{(0)}$ and extra pressure due to cell aggregation $\phi_n^{(0)}$.  In interstitial domain, the original ansatz (\ref{eqn:appform}) remains suitable.

\subsubsection{Microscale cell problems}

Substituting (\ref{eqn:newansatz1}), (\ref{eqn:newansatz2}) into the conservation of mass equations (\ref{eqn:3ddndtO1}) and (\ref{eqn:3ddwdtO1}), and the momentum equations (\ref{eqn:3dpnO2}) and (\ref{eqn:3dpwO2}), we obtain the following modified Stokes-type cell problems in $\Omega_{\Ti}$
\begin{align}
&\boldsymbol{\nabla}_{\boldsymbol{x}} \cdot \boldsymbol{K}_n^\T = \boldsymbol{0},\quad \boldsymbol{\nabla}_{\boldsymbol{x}} \cdot \boldsymbol{M}_n^\T = \boldsymbol{0},\label{eqn:StokesT1}\\
&\boldsymbol{\nabla}_{\boldsymbol{x}} \cdot \boldsymbol{K}_w^\T = \boldsymbol{0},\quad \boldsymbol{\nabla}_{\boldsymbol{x}} \cdot \boldsymbol{M}_w^\T = \boldsymbol{0},\label{eqn:StokesT2}\\
&\boldsymbol{\nabla}_{\boldsymbol{x}} \boldsymbol{a}_n^\T - \boldsymbol{I} - \nabla_{\boldsymbol{x}}^2 \boldsymbol{K}_n- \beta \theta^*_w\left(\boldsymbol{K}_w-\boldsymbol{K}_n\right)=\boldsymbol{0},\label{eqn:StokesT3}\\
&\boldsymbol{\nabla}_{\boldsymbol{x}} \boldsymbol{a}_w^\T - \boldsymbol{I} - \mu\nabla_{\boldsymbol{x}}^2 \boldsymbol{K}_w- \beta \theta^*_n\left(\boldsymbol{K}_n-\boldsymbol{K}_w\right)=\boldsymbol{0},\label{eqn:StokesT4}\\
&\boldsymbol{\nabla}_{\boldsymbol{x}} \boldsymbol{b}_n^\T - \boldsymbol{I} - \nabla_{\boldsymbol{x}}^2 \boldsymbol{M}_n- \beta \theta^*_w\left(\boldsymbol{M}_w-\boldsymbol{M}_n\right)=\boldsymbol{0},\label{eqn:StokesT5}\\
&\boldsymbol{\nabla}_{\boldsymbol{x}} \boldsymbol{b}_w^\T  - \mu\nabla_{\boldsymbol{x}}^2 \boldsymbol{M}_w- \beta \theta^*_n\left(\boldsymbol{M}_n-\boldsymbol{M}_w\right)=\boldsymbol{0}. \label{eqn:StokesT6}
\end{align}

Similarly, in $\Omega_{\Ii}$, standard Stokes problems are obtained, as follows:
\begin{align}
&\boldsymbol{\nabla}_{\boldsymbol{x}} \cdot \boldsymbol{K}_{\Ii}^\T = \boldsymbol{0},\\
&\boldsymbol{\nabla}_{\boldsymbol{x}} \boldsymbol{a}_{\Ii}^\T - \boldsymbol{I} - \mu\nabla_{\boldsymbol{x}}^2 \boldsymbol{K}_{\Ii}= \boldsymbol{0}.
\end{align} 

These cell problems are coupled together through the conditions (\ref{eqn:O1bc1})--(\ref{eqn:O1t}) and (\ref{eqn:oebc4}) specified on the interface, $\Gamma$, which supply
\begin{align}
& \boldsymbol{K}_{\Ii}^\T\bnn= \boldsymbol{0},\quad \boldsymbol{K}_n^\T \bnn= \boldsymbol{0},\quad\boldsymbol{K}_w^\T\bnn= \boldsymbol{0}, \quad \boldsymbol{M}_n^\T \bnn= \boldsymbol{0},\quad\boldsymbol{M}_w^\T\bnn= \boldsymbol{0},\\
&-\boldsymbol{a}_{\Ti}\otimes\bld{n}+\left(\bn\boldsymbol{K}_{\Ti}+\left(\bn\boldsymbol{K}_{\Ti}\right)^\T \right)\bld{n}=-\boldsymbol{a}_{\Ii}\otimes\bld{n}+\mu\left(\bn\boldsymbol{K}_{\Ii}+\left(\bn\boldsymbol{K}_{\Ii}\right)^\T\right)\bld{n},\\
&-\boldsymbol{b}_{\Ti}\otimes\bld{n}+\left(\bn\boldsymbol{M}_{\Ti}+\left(\bn\boldsymbol{M}_{\Ti}\right)^\T\right)\bld{n}=\theta_n^*\left[-\boldsymbol{a}_{\Ii}\otimes\bld{n}+\mu\left(\bn\boldsymbol{K}_{\Ii}+\left(\bn\boldsymbol{K}_{\Ii}\right)^\T\right)\bld{n}\right],\label{eq:stokesBC}
\end{align}
in which (\ref{O1paremacroscale}) has been employed to replace $p^{(0)}$, and where
\begin{align}
&\boldsymbol{K}_{\Ti}=\theta_n^*\boldsymbol{K}_n+\mu \theta_w^*\boldsymbol{K}_w, \quad \quad \boldsymbol{a}_{\Ti}=\theta_n^*\boldsymbol{a}_n+\theta_w^*\boldsymbol{a}_w,\\
&\boldsymbol{M}_{\Ti}=\theta_n^*\boldsymbol{M}_n+\mu \theta_w^*\boldsymbol{M}_w, \quad \quad \boldsymbol{b}_{\Ti}=\theta_n^*\boldsymbol{b}_n+\theta_w^*\boldsymbol{b}_w.
\end{align}
Lastly, on $\p \Omega_{\Si}$, (\ref{eqn:bcos2c}) provides
\be
\boldsymbol{K}_n= \boldsymbol{0},\quad\boldsymbol{K}_w= \boldsymbol{0},\quad \boldsymbol{M}_n= \boldsymbol{0},\quad\boldsymbol{M}_w= \boldsymbol{0}.
\ee
For uniqueness in the above cell problems, we use a standard approach see, \textit{e.g.}~\cite{mei2010homogenization,odea2015multiscale,penta2014effective,shipley2010multiscale}) and impose that in the relevant domain
\be
	\langle \boldsymbol{a}_{i} \rangle =  \boldsymbol{0},\quad \langle \boldsymbol{b}_{i} \rangle =  \boldsymbol{0}.
\ee

We note that, while a standard Stokes-type cell problem is obtained in  $\Omega_{\Ii}$, the multiphase dynamics in 
$\Omega_{\Ti}$ leads to signifcantly increased complexity. In particular, we obtain a set of coupled modified Stokes problems, determining the permeability tensors $\bK_{i}$, $\bM_{i}$ and extra pressures $\boldsymbol{a}_{i}$, $\boldsymbol{b}_{i}$ for each phase, which are further coupled to the flow in $\Omega_{\Ii}$ via stress and velocity continuity boundary conditions. Furthermore, we highlight that whilst the number of cell problems has increased as a result of the change in ansatz from that employed in \cite{holden2018multiphase}, we find that the permeability tensors are no longer dependent on macroscale pressures. The system we obtain therefore represents a significant simplification, taking the more familiar de-coupled form, whereby the quasi-steady cell problems can be solved separately from the macroscale description, that we obtain below.

\subsubsection{Averaging}

The macroscale flow is obtained by averaging (\ref{eqn:newansatz2}) via the definition (\ref{eq:av}) to obtain
\be
\langle \bv_{i}^{(0)} \rangle =-\langle \bK_{i}\rangle \bn_{\bX} p_w^{(0)}-\langle\bM_{i}\rangle \bn_{\bX} \phi_n^{(0)},\label{eqn:macroflow}
\ee
wherein $p_w{(0)}$ and $\phi_n^{(0)}$ (via $\theta_n^{(0)}$, and equations (\ref{eq:phi_n}), (\ref{eq:linfuncs1})) are obtained from the following system, derived from the average (exploiting the divergence theorem) of equations (\ref{eqn:3ddndtO2}), (\ref{eqn:oeT2}), (\ref{eqn:dcdt1}) and (\ref{eqn:dcdt2}):
\begin{align}
&\label{eqn:macron}
\frac{\partial }{\partial t}\langle \theta_n^{(0)}\rangle_{\Ti}+\theta_n^*\left(\bn_{\bX}\cdot\langle\bv_{n}^{(0)}\rangle_{\Ti}+\langle\bv_{\Gi}^{(1)}\cdot\bnn\rangle_{\Gi}\right)=\langle S^{(0)}_{n}\rangle_{\Ti},\\
&\label{eqn:macrop}
\bn_{\bX}\cdot\left(\tilde{\bK}\bn_{\bX} p_w^{(0)}+\tilde{\bM}\bn_{\bX} \phi_n^{(0)}\right)=-\langle\left(1-\bar{\rho}\right)S^{(0)}_{n}\rangle_{\Ti},\\
&\label{eqn:macroc}
\Phi_{\Ti\cup\Ii}\frac{\partial c^{(0)}}{\partial t}+c^*\left(1-\bar{\rho}\right) \langle S_{n}^{(0)}\rangle_{\Ti} =-\langle\Lambda^{(0)}\rangle_{\Ti}.
\end{align}
Equation (\ref{eqn:macroc}) arises from the sum of the averaged form of (\ref{eqn:dcdt1}) and (\ref{eqn:dcdt2}) and the tensors $\tilde{\bK}$ and $\tilde{\bM}$ are given by
\be
\tilde{\bK}=\langle\theta_n^*\bK_{n}+\theta_w^*\bK_{w}\rangle_{\Ti}+\langle\bK_{\Ii}\rangle_{\Ii},\quad \tilde{\bM}= \langle\theta_n^*\bM_{n}+\theta_w^*\bM_{w}\rangle_{\Ti}+\langle\theta_n^*\bK_{\Ii}\rangle_{\Ii},
\ee
where the individual permeability tensors $\bK_{i}$ and $\bM_{i}$ are determined from the set of coupled Stokes problems (\ref{eqn:StokesT1})--(\ref{eq:stokesBC}).

We remark that while the modification to the unit cell problems outlined above is significant, the impact of our modification to the approach of \cite{holden2018multiphase} on the macroscale description is less significant, being restricted to the redefinition of the relevant permeability tensors, and the associated velocities and pressures (in particular in the the explicit appearance of $\bn_{\bX}\phi_n^{(0)}$ terms associated with active cell motion). The governing system itself is of identical structure, and comprises a macroscale Darcy flow PDE, coupled to reaction equations describing tissue component volume fractions and nutrient concentration. Lastly, we note that as is common in analyses of this type, the macroscale model we obtain is not closed: we are required to specify constitutively the O$(\varepsilon)$ boundary velocity $\bv^{(1)}_{\Gi} \cdot \boldsymbol{n}$ (\textit{cf.}~\cite{irons2017microstructural,collis2016transport}). This is explored in \cite{holden2018multiphase} by means of detailed investigation of the travelling wave properties of the microscale multiphase model, but we do not pursue this here.

Lastly, we note in passing that in the limit case of inviscid water (that employed in \cite{holden2018multiphase} for illustrative numerical simulations), the overall pressure $p^{(0)}$ is zero and consequently so is $p_{w}^{(0)}+\theta_n^*\phi_{n}^{(0)}$. This means that the new ansatz (\ref{eqn:newansatz1}), (\ref{eqn:newansatz2}) can be rewritten as
\begin{align}
&\bv_{i}^{(0)}=\left[-\bK_{i} + \frac{1}{\theta^*_n}\bM_{i}\right]\bn_{\bX} p_w^{(0)},\nonumber\\
&p_{i}^{(1)}=\left[-\boldsymbol{a}_{i} + \frac{1}{\theta^*_n}\boldsymbol{b}_{i} \right]\bn_{\bX} p_w^{(0)} -\bar{p}_{i},\label{eqn:newansatzinviscid}
\end{align}
which is equivalent to the form used in \cite{holden2018multiphase} (see equation (3.17) therein) where the terms in square brackets are given by single tensors.

\section{Alternative boundary conditions}

\subsection{Cell motion on the scaffold surface}

In the model described above, we impose no-slip and no-penetration conditions on the scaffold boundary $\Omega_\Si$. While these are a sensible choice, reflecting the solid nature of the scaffold material, in some cases, a less restrictive choice may be of interest. For example, as well as the active motion embodied by the intraphase pressure $\phi_n$, cells may exhibit significant haptotactic motion on the scaffold surface itself. This is especially pertinent to the tissue engineering application under study, in which scaffolds may be produced to include substrate-bound chemoattractants thereby promoting cell ingress (see, \textit{e.g.}, \cite{wen2015haptotaxis,miller2006dose,miller2011spatially} and references therein). 

We do not consider a haptotactic model here, but consider the following simple alternative choice of boundary condition permitting cell motion
\be
\bv_n = b \frac{\partial\bv_n}{\partial n}\label{eq:slip}
\ee
where $b$ is a constant of proportionality and $\partial/\partial n$ denotes the normal derivative. We retain the no-penetration condition $\boldsymbol{v}_i\cdot\boldsymbol{n}=0$ on $\partial \Omega_\Si$ since the scaffold remains solid.

The effective macroscale equations remain the same in each case and the only change to the Stokes problem is in the tissue-scaffold boundary conditions. The above equations give
\be
\bK_n = b \frac{\partial \bK_n}{\partial n }, \quad \bM_n = b \frac{\partial \bM_n}{\partial n }, \quad \bK_n^\T\bnn=\bo, \quad \bM_n^\T\bnn=\bo
\ee
as the set of alternative boundary conditions to be applied on $\bK_n$ and $\bM_n$ at $\partial\Omega_{\Si}$ in the Stokes problem. (Note that the slip condition is of similar form to that obtained by Irons \emph{et al.} \cite{irons2017microstructural} in a similar cell problem, for a porous medium growth model.)

\subsection{Nutrient flux}

For completeness, we also indicate the influence of alternative concentration flux boundary conditions applied on the tissue-interstitium interface $\Gamma$. In the model developed above, we impose continuity of flux and concentration.  As discussed in \cite{shipley2010multiscale} two further options are suggested, which we now consider:

\paragraph{Option 1 -- Membrane law}
The flux of nutrient concentration across the boundary is proportional to the concentration jump. This widely-used approach demands:
\be
\left(c_\Ii\bv_\Ii-D_\Ii\bn c_\Ii\right)\cdot\bnn=\left(c_\Ti\bv_\Ti-D_\Ti\bn c_\Ti\right)\cdot\bnn=r\left(c_\Ti-c_\Ii\right),
\label{eq:membranebc}
\ee
where $r$ is a constant reflecting the permeability of the tissue boundary to nutrient flux.

\paragraph{Option 2 -- Concentration jump due to species solubility}
Alternatively, a concentration jump may be permitted, as a consequence of reduced solvability in the tissue compared to the interstitium (\textit{cf.} Henry's law for gases in which the concentration $c$ and partial pressure $P$ of a gas in solution are related through the $c=\gamma P$, where $\gamma$ denotes the solvability):
\be
\alpha c_\Ii=c_\Ti,
\label{eq:jumpbc}
\ee
where we assume for simplicity that $\alpha$ is a constant, although in a more general formulation it may be suitable to specify $\alpha=\alpha(\theta_n)$.

In the following, we investigate the choice of boundary condition, and scaling of associated constant, on the effective macroscale description. Note that the choice of condition has no direct impact on the Stokes problem on the periodic cell.

\subsubsection{Option 1 -- Membrane law}

Firstly we linearise the boundary condition (\ref{eq:membranebc}); assuming that at steady state the nutrient concentration is equal and uniform across both domains, $\Omega_\Ii$ and $\Omega_\Ti$ we obtain:
\be
c^*\left(\bv_\Ii-\bv_\Gi-\frac{1}{Pe_\Ii}\bn c_\Ii\right)\cdot\bnn=c^*\left(\bv_\Ti-\bv_\Gi-\frac{1}{Pe_\Ti}\bn c_\Ti\right)\cdot\bnn=r\left(c_\Ti-c_\Ii\right).
\ee
In the subsequent multiple scales analysis we consider two further scaling subcases on the membrane permeability; namely $r=\mbox{O}(1)$ or $r=\varepsilon\bar{r}$, with $\bar{r}=\mbox{O}(1)$. At leading order the boundary condition reads in each case:
 \be
 -\frac{1}{Pe_\Ii}\bn_{\bx} c_\Ii^{(0)}\cdot\bnn =  -\frac{1}{Pe_\Ti}\bn_{\bx} c_\Ti^{(0)}\cdot\bnn = 
  \begin{cases} 
   r \left(c_\Ti^{(0)}-c_\Ii^{(0)}\right) \\
  0
  \end{cases}
 \ee
 
We recall that the leading order problem is quasi-steady, so there is no growth of $\Omega_{\Ti}$, flux of fluid across the interface or nutrient uptake (see \S\ref{sec:eqsateachorder}); correspondingly, and in line with the linearised model set-up, it is sensible to assume that there is no induced diffusive transport of nutrient across $\Gamma$ either. In the first sub-case this implies that, since $r\neq0$, $c_\Ti^{(0)}=c_\Ii^{(0)}$, whilst in the second sub-case there is already no interfacial transport at this order. In both cases, the leading order concentration in each domain is independent of $\bx$, and in the first only we require $c_\Ti^{(0)}=c_\Ii^{(0)}$. 

Following through the rest of the analysis as described above and in \cite{holden2018multiphase} for O$(1)$ membrane permeability, we find that the effective macroscale equation is unchanged,
 \be
\Phi_{\Ti\cup\Ii}\frac{\partial c^{(0)}}{\partial t} + c^*\left(1-\bar{\rho}\right)\langle S_n^{(0)} \rangle_\Ti = - \langle \Lambda^{(0)} \rangle_\Ti.
 \ee
 
In the second sub-case, there are minor differences imbued by the fact that a leading-order concentration jump may be permitted and we obtain the following macroscale representation in each domain
 \begin{align}
\Phi_{\Ti}\frac{\partial c_\Ti^{(0)}}{\partial t} + c^*\left(1-\bar{\rho}\right)\langle S_n^{(0)} \rangle_\Ti &=-\langle \bar{r}\left(c_\Ti^{(0)}-c_\Ii^{(0)}\right)\rangle_\Gi - \langle \Lambda^{(0)} \rangle_\Ti\\
   \Phi_{\Ii}\frac{\partial c_\Ii^{(0)}}{\partial t} &= \langle \bar{r}\left(c_\Ti^{(0)}-c_\Ii^{(0)}\right)\rangle_\Gi ,
  \end{align}
which are identical to those presented in \cite{shipley2010multiscale}, except that advective transport is linearised in our description. Macroscale nutrient concentration in this case is given by two coupled equations, one for each of $c_\Ti^{(0)}$ and $c_\Ii^{(0)}$.

 \subsubsection{Option 2 -- Concentration jump due to species solubility}
 We remark that when linearising the model equations we can no longer assume that at steady state nutrient concentration $c^*$ is uniform across the entire unit cell (unless $\alpha=1$, which returns us our original representation). We instead suppose that nutrient concentration is uniform in each domain, connected by the boundary condition, i.e.
\begin{align}
c_\Ti &= c^*_\Ti + \delta c_{\Ti 1} + ...\\
c_\Ii &= c^*_\Ii + \delta c_{\Ii 1} + ...
\end{align}
where 
 \be
\alpha c^*_\Ii=c^*_\Ti.\label{eqn:cca}
\ee
As previously, $c^*_\Ti$ is defined by (\ref{eq:steadystate}). 

The linearised and rescaled equations for the nutrient concentration are given by:
\begin{align}
&\varepsilon\frac{\partial c_{\Ti,1}}{\partial t}+\bn\cdot\left(c_\Ti^*\left(\theta_n^*\bv_{n,1}+\theta_w^*\bv_{w,1}\right)\right)=\frac{1}{Pe_\Ti}\nabla^2 c_{\Ti,1}-\Lambda_1 \mbox{   in }\Omega_\Ti,\\
&\varepsilon\frac{\partial c_{\Ii,1}}{\partial t}+\bn\cdot\left(c_\Ii^* \bv_{\Ii,1}\right)=\frac{1}{Pe_\Ii}\bn^2 c_{\Ii,1}\mbox{   in }\Omega_\Ii,\\
&\left[c_i^* \left(\bv_{i,1}-\bv_{\Gi,1}\right)\cdot\bnn-\frac{1}{Pe_i}\bn c_{i,1}\cdot\bnn\right]^+_-=0,\\
&\alpha c_{\Ii,1}=c_{\Ti,1} \mbox{   on }\Gamma.\label{eqn:lobcss}
\end{align}
All other equations remain unchanged from the original analysis, except that in cell proliferation and nutrient uptake terms $c^*$ is replaced by $c^*_\Ti$. In the folllowing, the subscripts associated with the linearisation are omitted for clarity.

At leading order, we find, via standard arguments, that both $c^{(0)}_{\Ii}$ and $c^{(0)}_{\Ti}$ are independent of the microscale and related by the leading order version of (\ref{eqn:lobcss}).

At O$(\varepsilon)$ the relevant equations are:
\begin{align}
&\frac{\partial c^{(0)}_{\Ti}}{\partial t}+c_\Ti^*\left(\bn_{\bx}\cdot\left(\theta_n^*\bv_{n}^{(1)}+\theta_w^*\bv_{w}^{(1)}\right)+\bn_{\bX} \cdot\left(\theta_n^*\bv_{n}^{(0)}+\theta_w^*\bv_{w}^{(0)}\right)\right)\nonumber\\
&\hspace*{1cm}=\frac{1}{Pe_\Ti}\nabla_{\bx}^2 c^{(1)}_{\Ti}-\Lambda^{(0)} \mbox{   in }\Omega_\Ti,\label{eqn:oectss}\\
&\frac{\partial c^{(0)}_{\Ii}}{\partial t}+c_\Ii^* \left(\bn_{\bx}\cdot \bv_{\Ii}^{(1)}+\bn_{\bX}\cdot \bv_{\Ii}^{(0)}\right)=\frac{1}{Pe_\Ii}\bn^2_{\bx} c^{(1)}_{\Ii}\mbox{   in }\Omega_\Ii,\label{eqn:oeciss}\\
&\left[c_i^* \left(\bv_{i}^{(1)}-\bv_{\Gi}^{(1)}\right)\cdot\bnn-\frac{1}{Pe_i}\left(\bn_{\bx} c^{(1)}_{i}+\bn_{\bX} c^{(0)}_{i}\right)\cdot\bnn\right]^+_-=0,\label{eqn:oebc1ss}\\
&\alpha c^{(1)}_{\Ii}=c^{(1)}_{\Ti} \mbox{   on }\Gamma.
\end{align}

On averaging (\ref{eqn:oectss}) and (\ref{eqn:oeciss}) over their domains, we obtain:
 \begin{align}
\Phi_{\Ti}\frac{\partial c_\Ti^{(0)}}{\partial t} + c_{\Ti}^*\left(1-\bar{\rho}\right)\langle S_n^{(0)} \rangle_\Ti &=\langle \frac{1}{Pe_{\Ti}}\bn_{\bx} c^{(1)}_{\Ti}\cdot\bnn \rangle_\Gi - \langle \Lambda^{(0)} \rangle_\Ti ,\label{eqn:macroctss}\\
  \Phi_{\Ii}\frac{\partial c_\Ii^{(0)}}{\partial t} &= -\langle \frac{1}{Pe_{\Ii}}\bn_{\bx} c^{(1)}_{\Ii}\cdot\bnn \rangle_\Gi.\label{eqn:macrociss}
  \end{align}
Averaging boundary condition (\ref{eqn:oebc1ss}) over $\Gamma$ and rearranging, we find that
\be
\langle \frac{1}{Pe_{\Ti}}\bn_{\bx} c^{(1)}_{\Ti}\cdot\bnn \rangle_\Gi-\langle \frac{1}{Pe_{\Ii}}\bn_{\bx} c^{(1)}_{\Ii}\cdot\bnn \rangle_\Gi = \left(c^*_{\Ti}-c^*_{\Ii}\right)\langle Q^{(0)}\rangle_\Gi,
\ee
where
\be
Q^{(0)}= \left(\bv_{\Ii}^{(1)}-\bv_{\Gi}^{(1)}\right)\cdot\bnn = \left(\bv_{\Ti}^{(1)}-\bv_{\Gi}^{(1)}\right)\cdot\bnn = \theta_w^*\left(\bv_{w}^{(1)}-\bv_{\Gi}^{(1)}\right)\cdot\bnn
\ee
describes the leading order flux of material across the boundary of the tissue domain.

Summing (\ref{eqn:macroctss}) and (\ref{eqn:macrociss}), and exploiting (\ref{eqn:cca}) and (\ref{eqn:lobcss}) to eliminate $c_{\Ii}^*$ and $c_{\Ii}^{(0)}$, we obtain
 \be
\left(\Phi_{\Ti}+\frac{1}{\alpha}\Phi_{\Ii}\right)\frac{\partial c_\Ti^{(0)}}{\partial t}  + c_{\Ti}^*\left(1-\bar{\rho}\right)\langle S_n^{(0)} \rangle_\Ti =c^*_{\Ti}\left(1-\frac{1}{\alpha}\right)\langle Q^{(0)}\rangle_\Gi - \langle \Lambda^{(0)} \rangle_\Ti ,
 \ee 
 and $Q^{(0)}$ must be determined. Note that when $\alpha=1$, i.e. we have continuity of concentration on the boundary, we obtain (\ref{eqn:macroc}) as in the original model, and $Q^{(0)}$ no longer appears. 
 

\section{Discussion}

In this paper, we have derived an effective description for a growing tissue, by means of two-scale asymptotics. We considered a rigid periodic lattice-like structure covered by a layer of growing tissue. The model is therefore applicable to problems in regenerative medicine, such as tissue growth within a tissue engineering scaffold (our primary motivation), or biofilm growth, for example in the subsurface or the fouling of filters. 

Multiscale homogenisation techniques are increasing in popularity in biologically-inspired models, with a recent series of studies seeking to incorporate growth \cite{collis2017effective,collis2016transport,odea2015multiscale,penta2014effective,holden2018multiphase}. As in \cite{holden2018multiphase}, here, we seek to accommodate a more complex description of tissue growth than one comprising a solid undergoing accretion \cite{odea2015multiscale,penta2014effective} or volumetric growth \cite{collis2017effective}, by employing a multiphase fluid tissue model that naturally accommodates aspects such as interstitial growth and active cell motion, while still obtaining a tractable macroscale description. (A multiphase approach was used in \cite{collis2016transport}; however, exploiting the limit of large interphase drag reduces the dynamics to effectively an accretion-type process.) In \cite{holden2018multiphase}, this deficiency was addressed to obtain an effective description of tissue growth that retains active cell motion is permitted, caused by their tendency to aggregate or repel. Analytical progress was effected by a linearisation that ameliorates problems associated with complex mass-transfer considered in the multiphase model; however, the macroscale description obtained was fully coupled to the microscale unit cell problems, thereby providing a significant computational challenge in the general case (decoupling is obtained in the inviscid limit case). Here, we address this feature by adopting a more suitable solution ansatz to describe the velocities and pressures in the system, that respects the linear structure of the relevant momentum equations. This analysis provides a macroscale model of very similar structure to that presented in \cite{holden2018multiphase}, parameterised by permeability tensors, provided by a set of modified Stokes-type cell problems. The contribution of this work is that, unlike that presented in \cite{holden2018multiphase}, the cell problems are independent of the macroscale description, leading to a system whereby the quasi-steady cell problems may be solved separately from the macroscale description, thereby greatly simplifying the computational difficulty associated with fully-coupled multiscale descriptions. Moreover, we also demonstrate how the model formulation is changed under a set of alternative microscale boundary conditions associated with, for example, cell motion over the scaffold surface, alternative nutrient flux dynamics across the tissue-interstitium boundary.

\bibliography{biblioabbr}
\end{document}